\documentclass[reprint,amsmath,amssymb,aps,prl]{revtex4-1}
\usepackage[dvips]{graphicx}
\usepackage{dcolumn}
\usepackage{bm}
\usepackage{float}
\usepackage{hyperref}
\usepackage{color}
\usepackage{cancel}
\usepackage{bmpsize}
\usepackage{amsmath}
\usepackage{amssymb}

\definecolor{linkcolor}{rgb}{0.9,0,0}
\definecolor{citecolor}{rgb}{0,0.6,0}
\definecolor{urlcolor}{rgb}{0,0,1}
\hypersetup{
	colorlinks, linkcolor={linkcolor},
	citecolor={citecolor}, urlcolor={urlcolor}
}

\newcommand{\bra}[1]{\left\langle #1\right|}
\newcommand{\ket}[1]{\left| #1\right\rangle}
\newcommand{\braket}[2]{\left\langle
	#1\vphantom{#2}\right|\left.#2\vphantom{#1}\right\rangle}
\newcommand{\ketbra}[2]{\left| #1\right\rangle\!\left\langle#2\right|}

\newcommand{\be}[0]{\begin{equation}}
\newcommand{\ee}[0]{\end{equation}}

\newcommand{\tr}[0]{{\rm Tr}}

\newcommand{\lra}\simeq

\graphicspath{{images/}}

\begin{document}

	\title{A scheme for quantum teleportation between discrete and continuous encodings of an optical qubit}
	
	\author{Alexander E. Ulanov$^{1,2}$, Demid Sychev$^{1}$, Anastasia A. Pushkina$^{1,2}$, Ilya A. Fedorov$^{1,3}$ and A. I. Lvovsky$^{1,3,4}$}
	\affiliation{$^1$Russian Quantum Center, 100 Novaya St., Skolkovo, Moscow 143025, Russia}
	\affiliation{$^2$Moscow Institute of Physics and Technology, 141700 Dolgoprudny, Russia}
	\affiliation{$^3$P. N. Lebedev Physics Institute, Leninskiy prospect 53, Moscow 119991, Russia}
	\affiliation{$^4$Institute for Quantum Science and Technology, University of Calgary, Calgary AB T2N 1N4, Canada}

	%
	\begin{abstract}

		Transfer of quantum information between physical systems of a different nature is a central matter in quantum technologies. Particularly challenging is the transfer between discrete- and continuous degrees of freedom of various harmonic oscillator systems.
		Here we implement a protocol for teleporting a continuous-variable optical qubit, encoded by means of coherent states, onto a discrete-variable, single-rail qubit --- a superposition of the vacuum- and single-photon optical states --- via a hybrid entangled resource. We test our protocol on coherent states of different phases and obtain the teleported qubit with a fidelity of $0.83\pm0.04$. We also find that the phase of the resulting qubit varies consistently with that of the input, having a standard deviation of $3.8^\circ$ from the target value. Our work opens up the way to wide application of quantum information processing techniques where discrete- and continuous-variable encodings are combined within the same optical circuit.
		
	\end{abstract}
	
	\maketitle
	\vspace{10 mm}
	

	\begin{figure}[b]
		\includegraphics[width=\columnwidth]{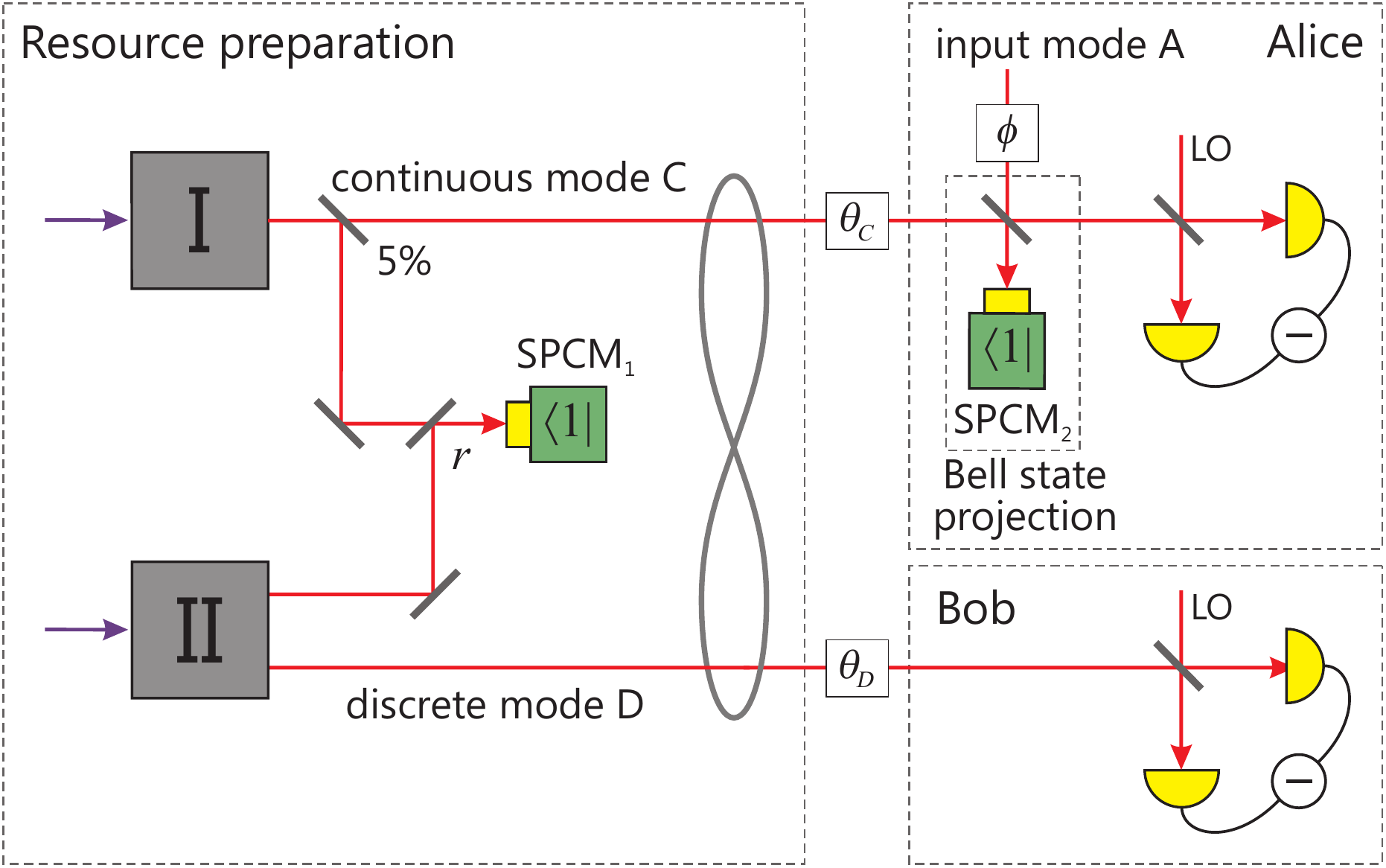}
		\caption{Conceptual scheme of the experiment. The teleportation occurs from a CV qubit (a coherent state in our case) in input mode A onto the DV vacuum-photon superposition in mode D. The preparation of the hybrid entangled resource (\ref{eq1}) in modes C and D is heralded by count of SPCM$_1$ while the Bell state projection is conditioned on a click of SPCM$_2$. The teleported state is measured by Bob using a homodyne detector. Another homodyne detector in mode C serves the auxiliary purpose of phase measurement.}
		\label{f1}
	\end{figure}
	
	Real-world application of quantum technologies in many cases requires simultaneous use of physical systems of a different nature \cite{Kur15} in the hybrid fashion. For example, one of the most promising platforms for quantum information processing is superconducting circuitry, best storage times are achieved in atomic systems, whereas transmission of that information in space is better realized with photons.
	
	Within quantum optics, the hybrid paradigm consists in symbiotic involvement of states and methods from discrete (DV) and continuous (CV) variable domains. Fortes of both can thereby be employed at once; for example, the continuous part takes advantage of the deterministic state preparation and relatively simple integration with existing information technologies \cite{BRA05}, while the discrete side benefits from a natural photodetection basis and ``built-in'' entanglement distillation after losses \cite{Kok07}. This flexibility allows hybrid quantum technologies to offer protocols which are superior to both pure CV and DV solutions \cite{Andersen15}. For example, using hybrid methods in quantum computing \cite{Qi2015} and error correction \cite{Lee13} is favorable, compared to the mainstream CV and DV protocols, in terms of the number of resources required. Long-distance distribution of CV entanglement requires hybrid processing to mitigate the effect of propagation losses \cite{Ulanov15, Ulanov16}.
	

	Taking advantage of these protocols however requires a procedure for transferring quantum data between continuous and discrete qubit systems \cite{Andersen15,Pirandola15}; 
	hybrid teleportation is therefore identified as one of the three key priorities for quantum teleportation science \cite{Pirandola2016}.
	In this work we make a step towards this goal, demonstrating quantum teleportation by means of a hybrid entangled channel. Specifically, we develop a technique to teleport a CV qubit, encoded as a superposition of coherent states, onto a DV qubit --- a superposition of the vacuum and single-photon Fock states. The protocol is conceptually different from the teleportation of photonic bits by a hybrid technique \cite{Takeda2013}, in which both the input and output states are discrete-variable single-rail qubits.
	
	\paragraph{Concept.}
	The protocol is carried out via entangled resource state \cite{Park2012,Lee13}
	\begin{align}
	\label{eq1}
	\ket{\textrm{R}}_{CD} &= \frac1{\sqrt2}\left(\ket{\alpha}_C\frac{\ket{0}_D+\ket{1}_D}{\sqrt 2} - \ket{-\alpha}_C\frac{\ket{0}_D-\ket{1}_D}{\sqrt 2}\right) \\ \label{eq1a}
	&= \frac12\left(\frac1{N_-}\ket{\mathrm{cat}_-}_C \ket{0}_D + \frac1{N_+}\ket{\mathrm{cat}_+}_C\ket{1}_D\right),
	\end{align}
	where $\ket{\mathrm{cat}_{\pm}}$ denote, respectively, positive and negative ``Schr\"odinger cat" states $N_\pm(\ket{\alpha}\pm\ket{-\alpha})$ \cite{our06,nee06,wak07} in the CV mode C and $\left\{\ket{0}_D, \ket{1}_D\right\}$ is the Fock basis for DV mode D, with $N_\pm=1/\sqrt{2\pm2e^{-2\alpha^2}}$ being the normalization factor.
	The CV part of the resource (\ref{eq1}) is distributed to Alice, while the DV part goes to Bob. The source state to be teleported is an amplitude-matched CV qubit
	\begin{align}
	\label{eqCVQ}
	\ket{\rm in}_A &=X\ket\alpha+Y\ket{-\alpha}\\ \nonumber
	&= \frac {X+Y}{2 N_+} \ket{\rm cat_+}_A +\frac {X-Y}{2 N_-} \ket{\rm cat_-}_A
	.
	\end{align}
	To implement the teleportation, Alice performs a joint measurement on this state and the CV part of the resource in the continuous-variable Bell basis defined by
	\begin{subequations}\label{eqBell0}
		\begin{align}
		\ket{\Phi^\pm}_{AC} &= \frac{\ket{\alpha}_A \ket{\alpha}_C \pm \ket{-\alpha}_A \ket{-\alpha}_C}{\sqrt{2\pm2e^{-4\alpha^2}}};\\
		\ket{\Psi^\pm}_{AC} &= \frac{\ket{\alpha}_A \ket{-\alpha}_C \pm \ket{-\alpha}_A \ket{\alpha}_C}{\sqrt{2\pm2e^{-4\alpha^2}}}.
		\end{align}
	\end{subequations}
	The Bell measurement implemented in our work, discussed below, approximates projection onto  state $\ket{\Phi^-}$, which can be rewritten as
	\begin{equation}\label{eqBell}
	\ket{\Phi^-}_{AC}=\frac{\ket{\rm cat_-}_A\ket{\rm cat_+}_C+\ket{\rm cat_+}_A\ket{\rm cat_-}_C}{\sqrt2}.
	\end{equation}
	In the event this state is detected in modes A and C, the state of the Bob's mode picks up the coefficients of the source CV qubit:
	\begin{align}
	\label{eqDVQ}
	\ket{\rm out}_D&=_{AC}\!\bra{\Phi^-}\left(\ket{\rm in}_A\otimes\ket{\rm \! R}_{CD}\right) \\ \nonumber
	&= \frac1{4N_+N_-}\left (\frac{X+Y}{\sqrt2}\ket{0}_D + \frac{X-Y}{\sqrt2} \ket{1}_D\right)
	\\ \nonumber
	&= \frac1{4N_+N_-}\left (X\frac{\ket{0}_D+\ket{1}_D}{\sqrt 2} + Y\frac{\ket{0}_D-\ket{1}_D}{\sqrt 2} \right)
	\end{align}
	completing teleportation of the quantum information between continuous and discrete bases. In particular, coherent states $\ket{\pm\alpha}$ are teleported onto $(\ket{0}_D\pm\ket{1}_D)/{\sqrt 2}$, states $\ket{\rm cat_+}$ and $\ket{\rm cat_-}$ onto $\ket 0$ and $\ket 1$, respectively.

	For high amplitudes $\alpha\gg1$, coherent states $\ket\alpha$ and $\ket{-\alpha}$ are largely orthogonal to each other, so the normalization factors $N_+\approx N_-\approx1/\sqrt2$. In this case, the scenario described above reduces to canonical quantum teleportation as proposed in 1993 by Bennett {\it et al.} \cite{QT}. 
	Limited by the current experimental technology, we work in the regime of moderate amplitudes $\alpha\sim 1/2$. This facilitates the preparation of the entangled resource, as well as 
	projecting the CV part of the resource and Alice's mode onto a Bell state (\ref{eqBell}). 
	
	The latter task can in principle be realized by bringing modes $A$ and $C$ into interference on a symmetric beam splitter followed by a  photon number parity measurement \cite{Jeong2001, Park2012}. However, because a coherent parity measurement is hard to implement in an experiment, we employ an alternative approach, which is feasible at moderate values of $\alpha \lesssim 1/2$. It consists in replacing a parity measurement by a single-photon measurement in mode $A$ after the interference. A success of that measurement implies that the beam splitter input state has been given by Eq.~(\ref{eqBell}).
	
	To see this, we observe that a symmetric beamsplitter
	transforms Bell states \eqref{eqBell0} into tensor products of a positive or negative Schr\"{o}dinger cat state in one output mode and vacuum in the other \cite{Jeong2001, Park2012}:
	\begin{subequations}\label{eqBell2}
		\begin{align}
		\ket{\Phi^\pm}_{AC} &\to  \frac{\left(\ket{\sqrt{2}\alpha} \pm \ket{-\sqrt{2}\alpha}\right)_A\otimes\ket{0}_C}{\sqrt{2\pm2e^{-4\alpha^2}}} \\
		\ket{\Psi^\pm}_{AC} &\to \frac{\ket{0}_A \otimes \left(\ket{\sqrt{2}\alpha} \pm \ket{-\sqrt{2}\alpha}\right)_C}{\sqrt{2\pm2e^{-4\alpha^2}}}.
		\end{align}
	\end{subequations}
	A click of a single-photon detector in mode A cannot happen in response to states (\ref{eqBell2}b). Further, state $\ket{\Phi^+}$ transforms into a positive cat, whose Fock decomposition for a small $\alpha$ contains primarily the vacuum state. The negative cat obtained from  $\ket{\Phi^-}$, on the other hand, has the single-photon state as the leading term in its Fock decomposition and hence is much more likely to bring about a  click (\textit{Supplementary}).

	
	We tested the above protocol in a simplified manner, with the CV qubit replaced by a coherent state of a varied
	phase: $\ket{\rm in}_A=\ket{\alpha e^{i\phi}}_A$. In order to predict the effect of our protocol on this state, we expand it in the Fock basis up to the single-photon term
	\begin{equation}\label{incohphi}
	\ket{\alpha e^{i\phi}}\approx \ket 0+\ket 1e^{i\phi}\approx \frac{1+e^{i\phi}}2\ket\alpha+\frac{1-e^{i\phi}}2\ket{-\alpha}.
	\end{equation}	
	Relating this to Eqs.~(\ref{eqCVQ}) and (\ref{eqDVQ}), the teleported state becomes
	\begin{align}\label{outcohphi}
	\ket{\rm out}_D
	&=\frac{\ket{0}_D+\ket{1}_De^{i\phi}}{\sqrt 2}.
	\end{align}	
	In other words, the protocol maps the phase of the input coherent state onto the azimuthal angle of Bob's qubit on the Bloch sphere. However, because state $\ket{\alpha e^{i\phi}}$ falls outside the qubit space spanned by $\{\ket\alpha,\ket{-\alpha}\}$ and hence decomposition \eqref{incohphi} holds only approximately, the teleportation fidelity will depend on the input phase $\phi$. For $\phi=0$ and $\phi=\pi$, we respectively have $\ket{\alpha e^{i\phi}}=\ket{\pm\alpha}$, so result \eqref{outcohphi} is expected to hold exactly, but  for other input phases the  fidelity is reduced.
	
	\begin{figure}[t]
		\includegraphics[width=\columnwidth]{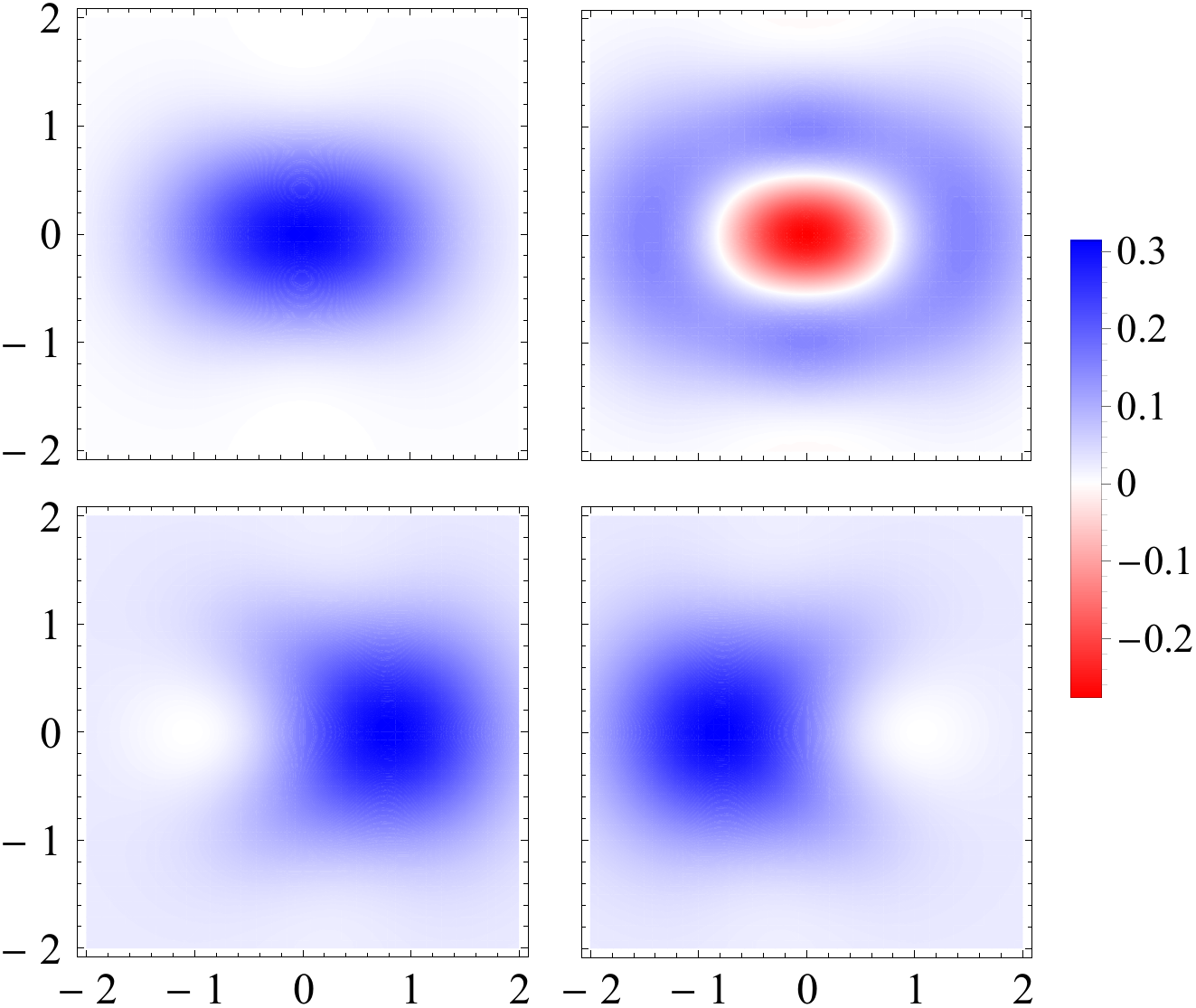}
		\caption{Top panels: Wigner functions of the CV mode of the entangled resource state \eqref{eq1}, projected in the DV mode onto $\ket{1}_D $ (left) and $ \ket{0}_D$ (right). Bottom panels: projections onto $(\ket{0}_D \pm \ket{1}_D)/\sqrt2$.}
		\label{f2}
	\end{figure}

	\paragraph{Experiment.}
	We produce the hybrid resource state (\ref{eq1}) following the method developed by Morin {\it et al.} \cite{Laurat2014} (Fig.~\ref{f1}). We employ two parametric down-conversion processes: one degenerate and the other non-degenerate, which take place in nonlinear crystals I and II. The crystals are periodically poled potassium titanyl-phosphate, and are pumped with frequency-doubled pulses at 390 nm, generated by a Ti:Sapphire laser with a repetition rate of 76 MHz and a pulse width of 1.5 ps \cite{instantFock}.

	The CV part of the resource is generated in crystal I, which initially produces a single-mode squeezed state with the squeezing parameter $\zeta=0.18$. This state approximates state $\ket{\rm cat_+}$ with $\alpha_+=\sqrt\zeta=0.42$ \cite{Our2009, Lvo2015}. The DV part of the resource originates from crystal II, where a weak two-mode squeezed state is created. The degree of squeezing in this state is much lower than that in the CV mode.
	
	In order to entangle the two parts of the resource, the CV mode is ``tapped" using a 5\% reflectivity beam splitter. The reflected mode is overlapped, on another beam splitter, with one of the modes emitted by crystal II and directed to a single-photon counting module SPCM$_1$. The resource is obtained conditioned on a click of that detector.

	\begin{figure*}[t]
		\includegraphics[width=\textwidth]{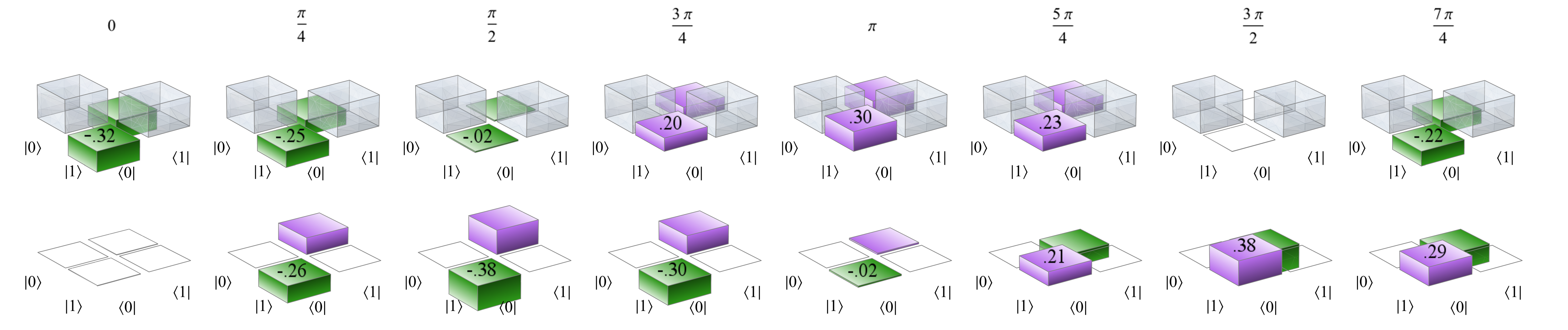}
		\caption{Density matrices of the teleported state, as measured by Bob's homodyne detector, for different phases of the input coherent state, reconstructed with a 54\% efficiency correction. Top row: real part. Bottom row: imaginary part. The coloring scheme distinguishes the diagonal (gray) as well as positive (purple) and negative (green) off-diagonal matrix elements for viewing convenience.}
		\label{DMs}
	\end{figure*}
	
	If the photon that causes the click comes from crystal I, then it is subtracted from the single-mode squeezed state, producing a negative Schr\"{o}dinger cat in the CV mode \cite{our06,Neer2011, nee06}. The DV mode is in the vacuum state in this case, because of the low degree of two-mode squeezing obtained from crystal II. If, on the other hand, the photon comes from crystal II, it heralds preparation of the single-photon state in the other output mode of that crystal \cite{Lvovsky2001}; the positive cat state in CV mode is then undisturbed. Because the detector is fundamentally unable to distinguish between these options, their coherent superposition, given by Eq.~(\ref{eq1}), is produced.
	
	We characterized the resource state by means of two-mode homodyne tomography. To that end, we temporarily removed the beam splitter associated with the Bell measurement and conducted a series of field quadrature measurements in modes C and D. The optical phases $\theta_{C,D}$ of the state in the two modes, required for the state reconstruction, were obtained from the quadrature data statistics. The data in the continuous channel, acquired without conditioning on an event in SPCM$_1$, corresponds to a squeezed vacuum state, so the quadrature variance is related to the phase according to $\langle X_C^2 \rangle \sim \cos 2\theta_C$, allowing us to extract $\theta_{C}$. Phase $\theta_D$ has been determined using correlation $\langle X_C X_D\rangle \propto \cos (\theta_D - \theta_C)$, which is characteristic of state (\ref{eq1}) as demonstrated in the \emph{Supplementary}. 
	
	The result of the maximum-likelihood reconstruction \cite{Lvo2004, Lvo2007}, with correction for 46\% of detection loss in both channels, has a fidelity of 96\% to state (\ref{eq1a}). 
	The top left panel of Fig.~\ref{f2} shows the Wigner function of the reconstructed state, projected onto $\ket{1}_D$ in the discrete mode, which is a positive cat state of amplitude $\alpha_+=0.42$. Projection onto $\ket{0}_D$, top right, corresponds to the photon-subtracted positive cat, which is a negative cat of amplitude $\alpha_-=0.67$, in agreement with theoretical expectation $\alpha_-=\sqrt{3}\alpha_+$ (\emph{Supplementary}). The results of projection onto $(\ket{0}_D \pm \ket{1}_D)/\sqrt2 $, which in the ideal case (\ref{eq1}) are coherent states, are shown in the bottom panels.
	The mismatch between $\alpha_+$ and $\alpha_-$ is an inherent shortcoming of the resource preparation protocol of Ref.~\cite{Laurat2014}. An alternative method \cite{Jeong14} does not have this drawback, but imposes much stronger limitations on the amplitude of the CV resource mode.

	The ratio of count rates $R_{\rm I,II}$ of SPCM$_1$ from the two crystals determines the relative weights of the components of the entangled resource. In order for them to be consistent with Eq.~(\ref{eq1a}), we must have 
	$$\frac{R_{\rm I}}{R_{\rm II}}=\left(\frac{N_+}{N_-}\right)^2=\frac{1-e^{-2\alpha_-^2}}{1+e^{-2\alpha_+^2}}\approx\frac13.$$
	Experimentally, this ratio is controlled by varying the reflectivity of the beam splitter in front of SPCM$_1$. The resulting resource preparation rate is $R_{\rm I} + R_{\rm II} \approx 20$ kHz.

	We perform the Bell projection on behalf of Alice by overlapping the input coherent state with the CV mode of the resource on a symmetric beamsplitter and subjecting one of the resulting modes to non-discriminating single-photon detection via SPCM$_2$. For the coherent state amplitudes used  in our work, state $\ket{\Phi^-}$ brings about a count event with the probability that is about ten times higher than the three other Bell states taken together (\textit{Supplementary}). Therefore such a count, in coincidence with a click from SPCM$_1$, heralds a successful event of the teleportation protocol. Such events occur in our experiment at a rate of 350 s$^{-1}$. The data collected in our experiment correspond to a total of 524288 events.

	\begin{figure}[b]
		\includegraphics[width=0.85\columnwidth]{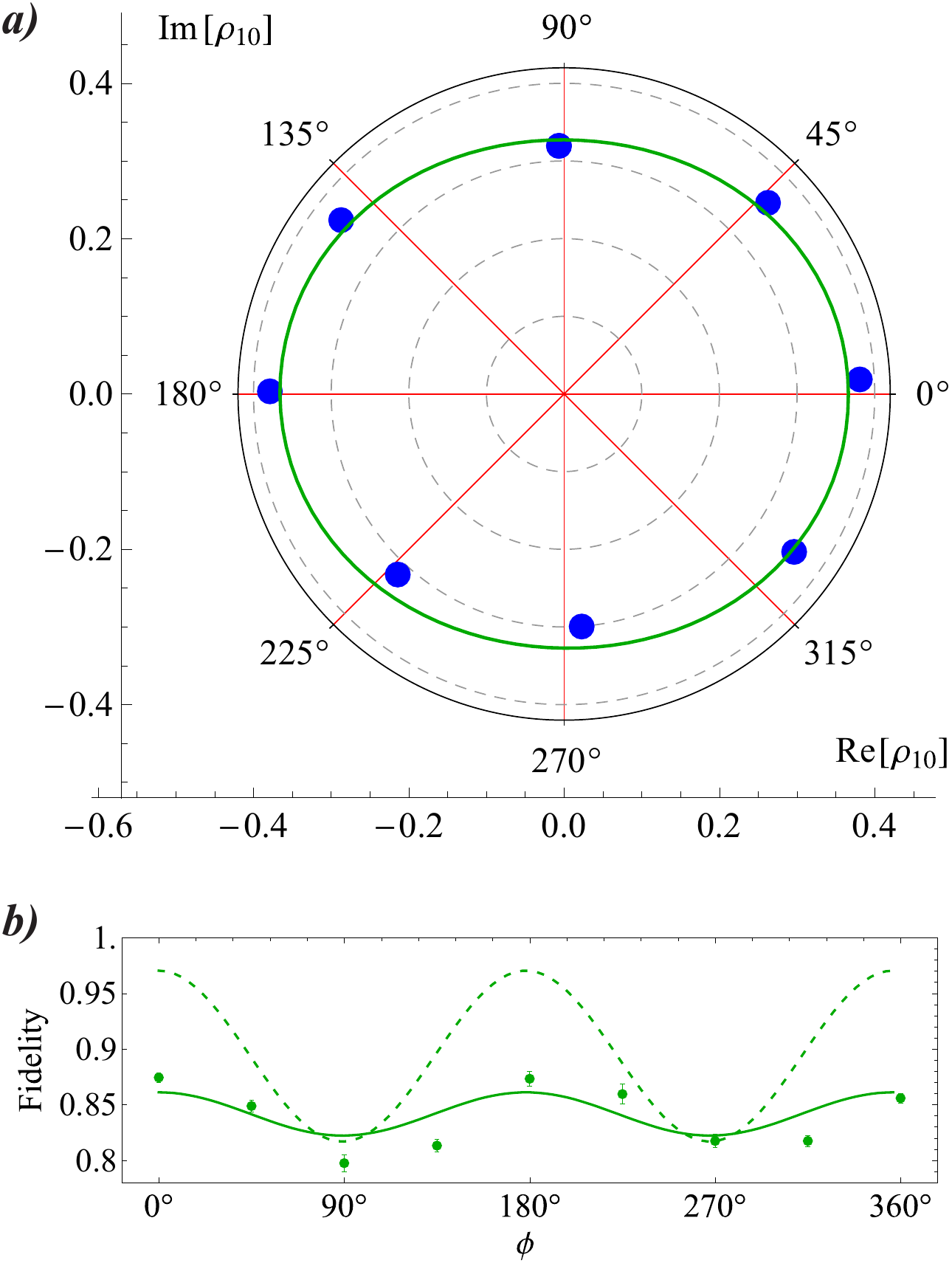}
		\caption{Top: off-diagonal elements of the teleported qubit's density matrix marked by blue dots. The red radial lines indicate the corresponding input state phases. Bottom panel: the fidelity between the states reconstructed from the experimental data with efficiency correction, and the theoretical expectation (\ref{outcohphi}). Error bars correspond to one standard deviation. The solid line is a theoretical prediction taking into account all experimental imperfections including the phase estimation uncertainty. The dashed line takes into account only the Bell measurement non-ideality.}
		\label{ReIm}
	\end{figure}
	
	To verify the teleportation and measure its fidelity, we subject the teleported state in mode D to optical homodyne tomography. The output state reconstruction and its comparison with the input state requires that phases $\theta_C,\theta_D,\phi$ be known at each moment in time. We obtained this information from the homodyne detector in mode C. Phases $\theta_C$ and $\theta_D$ of the resource modes are evaluated as discussed above, while the input coherent state phase $\phi$ is obtained from the average reading of Alice's homodyne detector: $\langle X_C \rangle \sim \cos \phi$. This phase, as well as the phase  $\theta_{D}$ of the discrete resource mode is varied by means of the piezo-electric transducer while the phase $\theta_{C}$ of the continuous mode is kept constant without active stabilization. In this way, for each value of $\phi$, a tomographically complete set of quadrature measurements is acquired.

	\paragraph{Results.}
	Figure \ref{DMs} shows the teleported qubit states for a set of coherent state phases $\phi$. The data are corrected for the 54\% detection efficiency on Bob's side.
	As expected from Eq.~\eqref{outcohphi}, the diagonal elements remain constant $\sim 1/2$, while the  off-diagonal elements change their argument dependent on $\phi$. This behavior is illustrated further in Fig.~\ref{ReIm}(a). The standard deviation between the phases of the input and teleported states is 0.067 rad or 3.8$^\circ$.

	The magnitude of the off-diagonal elements varies between the maximal value of 0.39, attained at real $\alpha$ ($\phi=0,\pi$) and the minimum of 0.27, which takes place at $\phi = \pm \pi/2$. This oscillatory effect is inherent to the application of the protocol to coherent states of different phases, as discussed above. It directly affects the teleportation fidelity as displayed in  Fig.~\ref{ReIm}(b).
	
	
	There are three additional effects leading to the degradation of the teleportation fidelity. The first one is the simplified Bell measurement protocol, whose fidelity falls  with increasing $\alpha$. Second, phase $\theta_C$, not stabilized during the experiment, fluctuates due to air movement with a mean deviation of 0.53 rad. The third detrimental factor is the $\sim 0.5$ rad statistical uncertainty of evaluating the resource phase difference $\theta_D - \theta_C$. This factor is present because the acquisition of the correlated quadrature data required for this evaluation is conditioned on clicks in SPCM$_1$, which occur at a relatively low rate of $\sim 20$ kHz. The combined effect of the two latter inefficiencies suppresses the oscillatory behavior and degrades the average fidelity by about 5\%, as shown by the dashed line in Fig.~\ref{ReIm}(b).   
	
	In order to apply our scheme to \emph{bona fide} CV qubits \cite{Lee13}, such qubits can be prepared using the technique of Ref.~\cite{Nielsen10}. Its integration with our experiment is straightforward, but requires significant enhancement of all nonlinear optical processes involved in order to bring up the proton count event rate.
	
	

\newpage
\setcounter{equation}{0}
\setcounter{figure}{0}
\makeatletter
\renewcommand{\thefigure}{S\@arabic\c@figure}
\makeatother
\section{Supplementary information}
\subsection{Bell measurement analysis}
We estimate the probabilities to obtain a count event of SPCM$_2$ in response to each of the four  Bell states in modes A and C. The Bell states are transformed by a symmetric beam splitter according to Eq.~(7) in the main text. Using the Fock state decomposition of the coherent state
\begin{equation}\label{}
\ket{\pm \sqrt2\alpha}=\sum_n e^{-\alpha^2}\frac{(\sqrt2\alpha)^n}{\sqrt{n!}}\ket n
\end{equation}
and the POVM element associated with a click of a non-discriminating single photon detector of efficiency $\eta$ 
\begin{equation}\label{POVM}
\hat \Pi_A=\sum_n [1-(1-\eta)^n]\ketbra nn,
\end{equation}
we estimate these probabilities for $\eta\ll 1$ to order $O(\alpha^8)$ as follows:
\begin{equation}\label{eqProbs}
p_{\ket{\Phi^+}} \approx 4\eta\alpha^4;\ 
p_{\ket{\Phi^-}} \approx\eta \left(1 + \frac43\alpha^4 \right);\  
p_{\ket{\Psi^\pm}}=0.	
\end{equation}

\begin{figure}[h]
	\includegraphics[width=0.8\columnwidth]{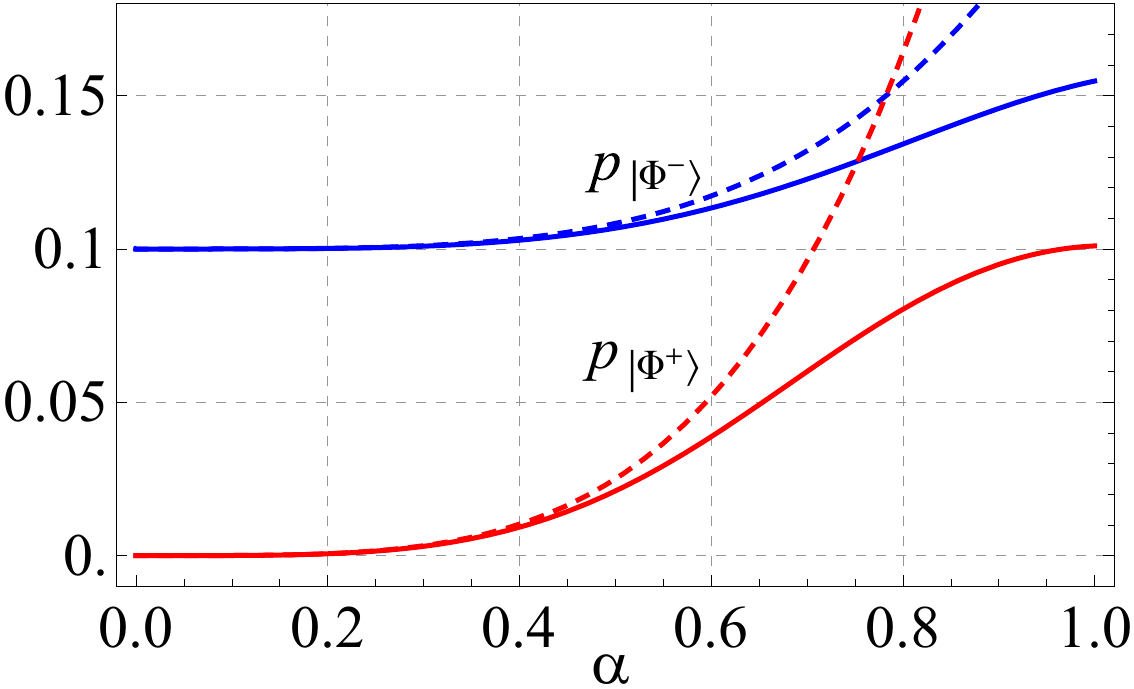}
	\caption{Bell state detection probabilities $p_{\ket{\Phi^+}}$ and $p_{\ket{\Phi^-}}$ for the experimental SPCM efficiency $\eta=0.1$. Solid lines show precise calculations, dashed lines  approximation (\ref{eqProbs}). }
	\label{s1}
\end{figure}

We see that, for small amplitudes, the ratio of these probabilities (Fig.~\ref{s1}) scales as $\alpha^4$, allowing for good fidelity in Bell state discrimination. In our experiment, we have $\alpha_+ = 0.42$ and $\alpha_-=0.67$, so $p_{\ket{\Phi^-}}/p_{\ket{\Phi^+}}=10.7$.

\subsection{Squeezed vacua vs. Schr\"odinger cats}
The momentum-squeezed vacuum state  with squeezing parameter $\zeta$ is decomposed into the Fock basis according to \cite{lvo15}
\begin{align}\label{nsq4}
\hat S(\zeta)\ket 0 &=\frac 1{\sqrt{\cosh \zeta}}\sum\limits_{m=0}^\infty(\tanh \zeta)^m\frac{\sqrt{(2m)!}}{2^m m!}\ket{2m}\\
&=\ket 0+\frac{1}{\sqrt{2}}\zeta\ket 2+\sqrt{\frac{3}{8}} \zeta
^2\ket 4+O(\zeta^3).\nonumber
\end{align}
Applying the photon annihilation operator to this state, we obtain
\begin{align}\label{nsq41}
\hat a\hat S(\zeta)\ket 0 
&=\zeta\left[\ket 1+\sqrt{\frac{3}{2}} \zeta
\ket 3+O(\zeta^2)\right],
\end{align}

On the other hand, using the Fock representation of a coherent state, we decompose the Schr\"odinger cat states $\ket{\mathrm{cat}_{\pm}}=\ket{\alpha}\pm\ket{-\alpha}$  as follows:
\begin{subequations}
	\begin{align}
	\ket{\rm cat_+}&=\ket 0+\frac 1{\sqrt 2}\alpha^2\ket{2}+O(\alpha^4);\\
	\ket{\rm cat_-}&=\ket 1+\frac 1{\sqrt 6}\alpha^2\ket{3}+O(\alpha^4).
	\end{align}
\end{subequations}

Comparing these results, we find that, for $\zeta\lesssim 1$ and up to a normalization factor,
\begin{itemize}
	\item a squeezed vacuum state $\hat S(\zeta)\ket 0$ approximates a positive Schr\"odinger cat state $\ket{\rm cat_+}$ with amplitude $\alpha=\sqrt\zeta$; 
	\item a photon subtracted squeezed vacuum state $\hat a\hat S(\zeta)\ket 0$ approximates a negative Schr\"odinger cat state $\ket{\rm cat_-}$ with amplitude $\alpha=\sqrt{3\zeta}$.	
\end{itemize}

\subsection{Quadrature correlation of the entangled resource state}
Here we find how phase shifts in the discrete and continuous modes of the entangled resource state affect the correlated quadrature measured by Alice's and Bob's homodyne detectors. Applying the phase shift operator $\hat U_{\theta_C\theta_D}=e^{i\hat n_C\theta_C+i\hat n_D\theta_D}$  to the resource state,  given by Eq.~(1) of the main text, we obtain
\begin{align*}\label{eq1s}
\hat U_{\theta_C\theta_D}\ket{\textrm{R}}_{CD} 
&= \ket{\alpha e^{i\theta_C}}_C\frac{\ket{0}_D+e^{i\theta_D}\ket{1}_D}{2} \\
&- \ket{-\alpha e^{i\theta_C}}_C\frac{\ket{0}_D-e^{i\theta_D}\ket{1}_D}{2} 
\end{align*}
Expressing the position operator as $\hat X=(\hat a+\hat a^\dag)/\sqrt2$ and recalling that the coherent state is an eigenstate of the annihilation operator, we find the expectation value of observable $\hat X_C\hat X_D$ in the phase shifted resource state
\begin{equation}\label{}
_{CD}\bra{\textrm{R}}\hat U^\dag_{\theta_C\theta_D} \hat X_C\hat X_D \hat U_{\theta_C\theta_D}\ket{\textrm{R}}_{CD} =\alpha\cos\left(\theta_C - \theta_D\right),
\end{equation}
which allows us to evaluate the phase difference $\theta_C-\theta_D$ as discussed in the main text.

\begin{figure}[t]
	\includegraphics[width=\columnwidth]{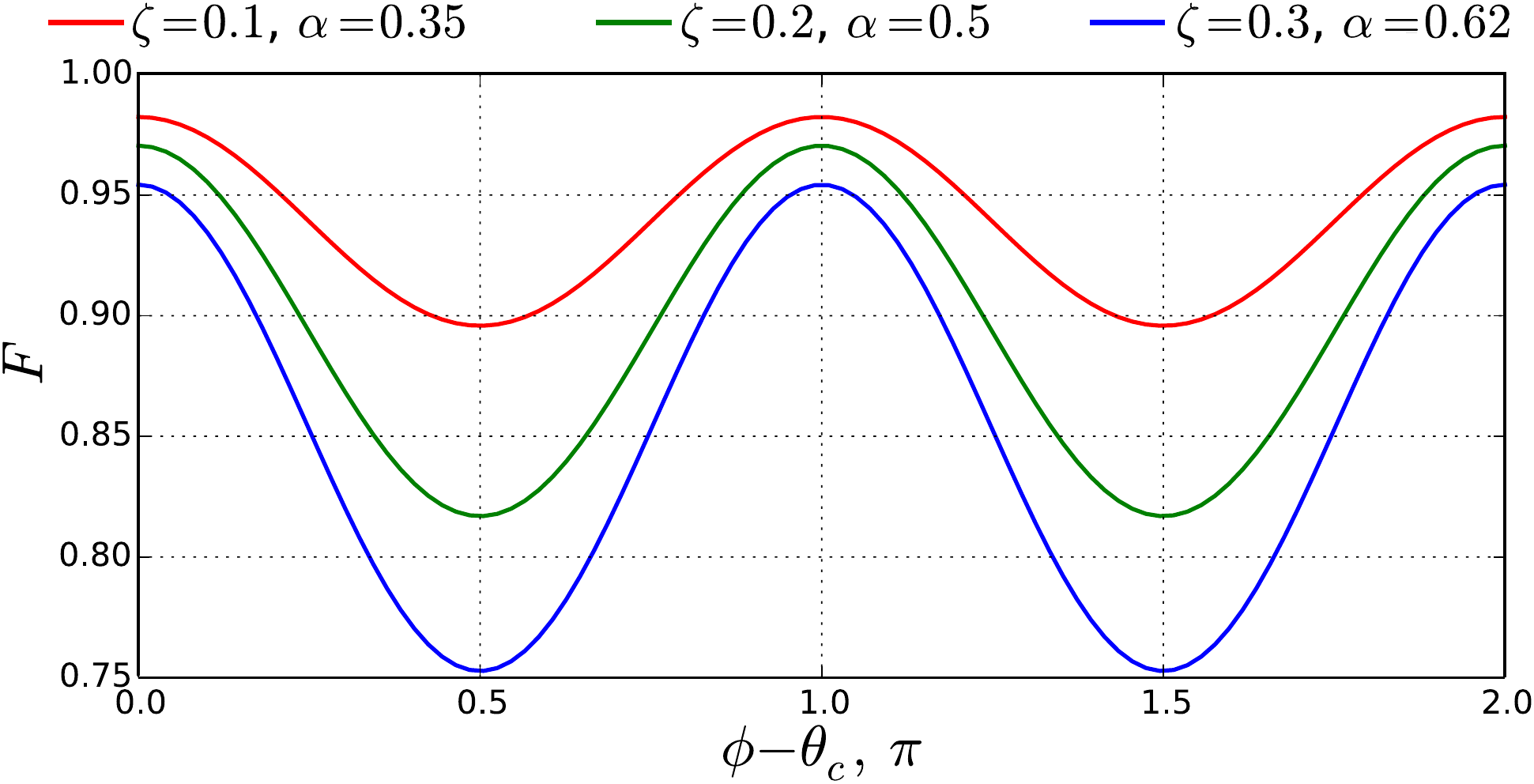}
	\caption{Fidelity of the teleported state with respect to ideal expectation (\ref{eqPure}) for different squeezing parameters $\zeta$ of the initial single--mode squeezed state. The corresponding amplitude $\alpha$ of the resource is indicated in the legend. Solid lines correspond to output state \eqref{eqDM} which accounts for an imperfect Bell measurement, with the solid green line corresponding to the dashed line in the Fig.3 main text. 
	}
	\label{s2}
\end{figure}

\begin{widetext} 
	
	\subsection{Teleportation of a coherent state}
	Here we analyze the application of the teleportation protocol to a coherent state $\ket{\alpha e^{i\phi}}$ in order to determine the output state and the fidelity. As in the previous section, we take into account the phases $\theta_C$ and $\theta_D$ of the two resource modes. The state of modes A, C and D prior to the Bell measurement  is given by
	\begin{align*}
	\ket{\alpha e^{i\phi}}_A \otimes \left[\hat U_{\theta_C\theta_D}\ket{\textrm{R}}_{CD}\right]&=
	\ket{\alpha e^{i\phi}}_A\ket{\alpha e^{i\theta_C}}_C\frac{\ket{0}_D+e^{i\theta_D}\ket{1}_D}{2} 
	-\ket{\alpha e^{i\phi}}_A\ket{-\alpha e^{i\theta_C}}_C\frac{\ket{0}_D-e^{i\theta_D}\ket{1}_D}{2}.
	\end{align*}
	Transforming modes A and C on the symmetric beam splitter state
	\begin{align*}
	\ket{\Psi_{\rm ACD}} = \ket{\alpha\frac{ e^{i\phi}+ e^{i\theta_C}}{\sqrt2}}_A\ket{\alpha\frac{ e^{i\phi}- e^{i\theta_C}}{\sqrt2}}_C\frac{\ket{0}_D+e^{i\theta_D}\ket{1}_D}{2} 
	-\ket{\alpha\frac{ e^{i\phi}- e^{i\theta_C}}{\sqrt2}}_A\ket{\alpha\frac{ e^{i\phi}+ e^{i\theta_C}}{\sqrt2}}_C\frac{\ket{0}_D-e^{i\theta_D}\ket{1}_D}{2}. 
	\end{align*}
	which for the resource phases $\theta_{C,D}$ fixed at zero gives Eq.~(8) in the main text.
	

	This state is subjected to single--photon detection in modes A and C, described by POVM $\hat{\Pi}_A\otimes\hat{\mathbb{I}}_C$, where $\hat{\Pi}_A$ is given by Eq.~\ref{POVM} and $\hat{\mathbb{I}}_C$ is the identity operator reflecting the absence of conditioning on a measurement result in mode C.
	The resulting state of mode D is
	\begin{equation}
	\label{Proj}
	\hat{\rho}_{D} = \tr_{AC} \left[\hat{\Pi}_A\otimes\hat{\mathbb{I}}_C \ket{\Psi_{\rm ACD}} \! \bra{\Psi_{\rm ACD}}\right],
	\end{equation}
	In the first order of $\alpha$, POVM $\hat{\Pi}_A\otimes\hat{\mathbb{I}}_C$ is approximated by projection onto one- and zero-photon states in modes A and C, respectively. The result of this projection is
	\begin{align}
	\label{eqP1}
	_{AC} \! \braket{10}{\Psi_{\rm ACD}}&= \alpha \dfrac{e^{i \phi} + e^{i \theta_C}}{\sqrt{2}} \times \frac{\ket{0}_D+e^{i\theta_D}\ket{1}_D}{2} 
	- \alpha \dfrac{e^{i \phi} - e^{i \theta_C}}{\sqrt{2}} \times \frac{\ket{0}_D-e^{i\theta_D}\ket{1}_D}{2} \\
	&\propto e^{i \theta_C} \ket{0}_D + e^{i \theta_D}e^{i \phi} \ket{1}_D,\label{eqPure}
	\end{align}
	which for the resource phases $\theta_{C,D}$ fixed at zero gives Eq.~(9) in the main text.
	
	In the second-order approximation of $\alpha$, the resulting state is no longer pure. In the small--efficiency limit $\eta \ll 1$ and after renormalization, Eq.~(\ref{Proj}) becomes
	\begin{equation}
	\label{eqDM}
	\hat{\rho}_D
	=
	\dfrac{1}{2}
	\left[ {\begin{array}{cc}
		1 & e^{-i \left(\phi + \theta_D - \theta_C\right)} \left[1+\alpha^2 (e^{2i(\phi-\theta_C)} -1)\right] \\
		e^{i \left(\phi + \theta_D - \theta_C\right)} \left[1+\alpha^2 (e^{-2i(\phi-\theta_C)} -1)\right] & 1 \\
		\end{array} } \right],
	\end{equation}
	which at small $\alpha$ reduces to the pure state (\ref{eqPure}).
	Relative to (\ref{eqPure}), state (\ref{eqDM}) has fidelity
	\begin{equation}
	\label{eqFid}
	F = 1-\frac{\alpha^2}2\{1-\cos[2(\phi-\theta_C)]\}.
	\end{equation}
	This result, shown in the Fig. \ref{s2}, exhibits the oscillatory behavior discussed in the main text. 
\end{widetext}

\end{document}